\documentclass[aps,amsfonts,pra,twocolumn,superscriptaddress,showpacs,floatfix]{revtex4}
\usepackage{epsfig,amsmath,amssymb,bm,epsf,graphics}

\def\bra#1{\langle#1\vert}
\def\ket#1{\vert#1\rangle}
\def\ketbra#1{\vert#1\rangle\langle#1\vert}
\def\ipr#1#2{\langle#1\vert#2\rangle}

\def\Longarrow{\protect\@lra}
\def\@lra{\relbar\joinrel\relbar\joinrel\relbar\joinrel%
          \relbar\joinrel\rightarrow}

\def\mygamma{r}
\begin{document}
\title{Synthesizing arbitrary two-photon polarization mixed states}
\author{Tzu-Chieh Wei}
\affiliation{Department of Physics,
University of Illinois at Urbana-Champaign,
Urbana, Illinois 61801-3080}
\author{Joseph B. Altepeter}
\affiliation{Department of Physics,
University of Illinois at Urbana-Champaign,
Urbana, Illinois 61801-3080}
\author{David Branning\footnote{Present address: Department of Physics and Optical Engineering, Rose-Hulman Institute of Technology, Terre Haute, IN 47803, USA.}}
\affiliation{Department of Physics,
University of Illinois at Urbana-Champaign,
Urbana, Illinois 61801-3080}
\author{Paul M. Goldbart}
\affiliation{Department of Physics,
University of Illinois at Urbana-Champaign,
Urbana, Illinois 61801-3080}
\author{D. F. V. James}
\affiliation{Theoretical Division, T-4, Los Alamos National Laboratory, Los Alamos, New Mexico 87545}
\author{Evan Jeffrey}
\affiliation{Department of Physics,
University of Illinois at Urbana-Champaign,
Urbana, Illinois 61801-3080}
\author{Paul G. Kwiat}
\affiliation{Department of Physics,
University of Illinois at Urbana-Champaign,
Urbana, Illinois 61801-3080}
\author{Swagatam Mukhopadhyay}
\affiliation{Department of Physics,
University of Illinois at Urbana-Champaign,
Urbana, Illinois 61801-3080}
\author{Nicholas A. Peters}
\affiliation{Department of Physics,
University of Illinois at Urbana-Champaign,
Urbana, Illinois 61801-3080}
\date{January 21, 2005}
\begin{abstract}
Two methods for creating arbitrary two-photon polarization {\it pure\/} states
are introduced. Based on these,
four schemes for creating two-photon polarization {\it mixed\/} states are proposed and
analyzed. The first two schemes can synthesize completely arbitrary 
two-qubit mixed states, i.e., control all 15 free parameters: Scheme~I
requires several sets of crystals, while Scheme~II requires only a single set,
but relies on decohering the pump beam. Additionally, we describe two further
schemes which are much easier to implement. Although the total capability of
these is still being studied, we show that they can synthesize all two-qubit
Werner states, maximally entangled mixed states, Collins-Gisin states,
and arbitrary Bell-diagonal states.
\end{abstract}
\pacs{03.67.Mn, 42.50.Dv, 42.65.Lm}

\maketitle

\section{Introduction}
\label{sec:intro}
Quantum information processing~\cite{NielsenChuang00} promises 
great power relative to its classical counterpart. 
Many quantum information processes
require specific pure entangled states, such as Bell states, to succeed.
After interacting with the environment, however, pure states inevitably decohere; decoherence generally causes pure entangled states to become mixed and less entangled. Quantum error correction~\cite{Shor95} 
and entanglement 
distillation/concentration~\cite{BennettBrassardPopescuSchumacherSmolinWootters96} have been developed to help cope with a noisy (and hence decohering) environment.
On the other hand, there are implementations
using mixed states to investigate quantum computing, e.g., 
liquid-state NMR~\cite{ChuangGershenfeldKubinec98}.
The states in this last example are highly mixed and have no entanglement.
Still, between highly entangled pure states and highly separable mixed
states there exists a vast experimentally unexplored region 
in Hilbert space (more precisely, the space of density matrices), 
where states can be simultaneously mixed and entangled. 
The two-qubit system possesses the simplest
and smallest Hilbert space that permits the existence of entanglement.
Separate from the specific protocols which make use of the states, it
is of fundamental interest to understand the preparation of one of
the most basic quantum systems.
Although there have been many attempts~\cite{White_James_Munro_Kwiat01,NickKwiat,ZhangHuangLiGuo02,BarbieriDeMartiniDiNepiMataloni03} to synthesize two-qubit mixed polarization states, 
none has yet been able to create completely arbitrary two-qubit mixed states~\cite{Zhang04}.

Here we describe several two-photon polarization state implementations  
that should in principle enable preparation of arbitrary two-qubit 
mixed states, including states possessing all physical degrees of
entanglement and entropy. 
The schemes we shall present facilitate state creation and
allow  access to two-qubit Hilbert space and can be useful
for current and future quantum information protocols. 
We remark that if there exist efficient two-qubit entangling gates such as
CNOT~\cite{NielsenChuang00}, arbitrary
state synthesis can be systematically implemented by first generating
a purification~\cite{Purification} of the mixed state by adding ancillas, 
and then
tracing over the ancillas. However, efficient photon-polarization CNOT 
gates do not exist~\cite{FootnoteCNOT}, so we rely on other degrees of freedom
 to introduce decoherence, leading to mixed states.

The paper is organized as follows. After a brief background
discussion in Sec.~\ref{sec:background},  we describe, in Sec.~\ref{sec:pure},
two schemes to achieve {\it arbitrary pure\/} two-photon polarization 
states by employing downconversion in a two-crystal arrangement.
The first one is based on the existence of Schmidt decompositions.
The second one utilizes the coherent superpositions of two downconversion
processes embedded in an interferometric setup. 
In Sec.~\ref{sec:Arbmixed} we describe how to extend these two schemes 
to realize
arbitrary two-qubit {\it mixed\/} states (Schemes I and II). 
In Sec.~\ref{sec:Redmixed} we propose two reduced schemes (III and IV) that provide
practical ways to realize several important families of states that 
are currently of interest, including Werner states~\cite{Werner89}, 
maximally entangled mixed states~\cite{Munro_James_White_Kwiat01}, 
Collins-Gisin states~\cite{CollinsGisin03}, and arbitrary Bell-diagonal states.
Scheme III requires only two downconversion crystals, but cannot 
synthesize all two-qubit states. Scheme IV partially extends the set of
attainable states, but requires four downconversion crystals.
Finally, in Sec.~\ref{sec:conclude} we summarize
the four schemes and mention possible applications. Readers who do not
require full details but want an overview of the four schemes
and synthesizable states can refer to Table~\ref{tbl:comparison}. The details
of how to create particular families of states can be found around the equations [(\ref{eqn:MEMS}), (\ref{eqn:Werner}), (\ref{eqn:CG}), and (\ref{eqn:BD})] describing
these states.
 \subsection{Background information}
\label{sec:background}
The entangled photon pairs we consider come from frequency-degenerate type-I
spontaneous parametric downconversion (SPDC)~\cite{Kwiat_Waks_White_Appelbaum_Eberhard99}.
The general state from SPDC is a two-mode squeezed state consisting
of vacuum and $k$-pair states~\cite{OuWangZouMandel90}:
\begin{equation}
\label{eqn:2MS}
\ket{\Psi}=\ket{\rm vacuum}+\varepsilon\ket{\psi^{(1)}}+\varepsilon^2\ket{\psi^{(2)}}
+\cdots,
\end{equation}
where $\ket{\psi^{(k)}}$ is a $k$-pair state, and $\varepsilon$ is the
relative amplitude (typically of order $10^{-6}$) to find a single pair. 
The post-selected 1-pair state $\ket{\psi^{(1)}}$ is composed of
two daughter photons, usually called {\it signal\/} and {\it idler\/}.  
For the present article, we limit our attention to the case where the
signal and idler photons have approximately degenerate central frequencies, 
half that of the pump. (Our schemes apply to nondegenerate case as well.)
When the downconversion momenta are well collimated or otherwise
sharply selected (experimentally by a small iris), one can neglect
the momentum dependence of the pair state.  
The post-selected two-photon state can then be described by  
\begin{equation}
|\psi^{(\!1\!)}(\omega)\rangle=
\underbrace{\Big\{\sum_{j,k}c_{jk}|\chi_{j},\chi_k\rangle\Big\}}_{\rm polarization}
\!\otimes\!\underbrace{\int_{\phantom{j_{{j_{j}}_j}}}\!\! d\epsilon\, 
A_{si}(\epsilon)\,|\frac{\omega}{2}\!+\!\epsilon,\frac{\omega}{2}\!-\!\epsilon\rangle}_{\rm frequency},
\label{eqn:psi0}
\end{equation}
where $\omega$ is the pump frequency.  $|\chi_j,\chi_k\rangle$ and $|\frac{\omega}{2}+\epsilon,\frac{\omega}{2}-\epsilon\rangle$ 
respectively represent the 
polarizations and frequencies of the two photons, with
$|\chi_{1}\rangle\equiv|H\rangle$, the horizontal polarization, and 
$|\chi_{2}\rangle\equiv|V\rangle$, the vertical polarization.  $c_{jk}$ is the
amplitude of the polarization state $|\chi_j,\chi_k\rangle$; for single-crystal type-I
phase-matching the polarization state is unentangled, i.e.,   $c_{jk}=a_j b_k$.
$A_{si}(\epsilon)$ 
is the amplitude for a particular division of energy, so that  
$\epsilon$ indicates the deviation from half pump frequency.
$|A_{si}(\epsilon)|^2$ is peaked at $\epsilon=0$ with
width $\delta_\epsilon$, and we shall approximate it 
by a gaussian distribution: 
\begin{equation}
\label{eqn:Asi}
|A_{si}(\epsilon)|^2=\frac{1}{\sqrt{2\pi\delta^2_\epsilon}}
\exp\left({-\frac{\epsilon^2}{2\delta^2_\epsilon}}\right).
\end{equation}

More generally, the pump is not monochromatic, 
and therefore the pair state should be described by
\begin{equation}
|\psi\rangle=\int\! d\omega A_p(\omega) |\psi^{(1)}(\omega)\rangle,
\label{eqn:psi00}
\end{equation}
where $A_p(\omega)$ describes the frequency spread of the pump, assumed to
be peaked at some frequency $\omega_0$ with half-width $\delta_\omega$. 
For most of the following discussion, we consider thicknesses
of waveplates and crystals that are much less than the coherence length
${l}_p$ ($\equiv c/\delta_\omega$) of the pump, 
and hence we can safely use Eq.~(\ref{eqn:psi0}). 
The coherence length of downconversion photons (${l}_{si}\equiv c/\delta_\epsilon$) is usually much smaller than ${l}_{p}$,
i.e., $c/\delta_\epsilon\ll c/\delta_\omega$, because 
there are many ways to distribute the energy of the pump photon between
the daughter photons in each pair, resulting in a large 
$\delta_\epsilon$~\cite{footnote:coherencelength}.

\section{Schemes for arbitrary two-photon polarization pure states}
\label{sec:pure}
\subsection{Via Schmidt decomposition}
\label{sec:pure1}
\begin{figure} 
\vspace{0.5cm}
\centerline{
\rotatebox{0}{
        \epsfxsize=8cm
        \epsfbox{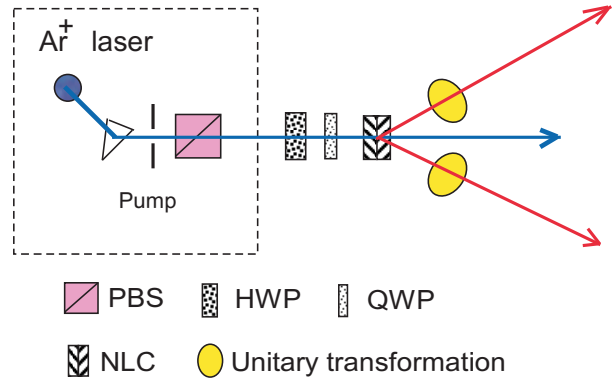}
}
}
\caption{(Color online) Arbitrary pure states via Schmidt decomposition. 
PBS: polarizing beam splitter; 
HWP: half-waveplate; 
QWP: quarter-waveplate; 
NLC: nonlinear crystals. }
\label{fig:ArbPureSch1}
\end{figure}

Using the method of Schmidt decomposition~\cite{NielsenChuang00}, an arbitrary 
two-qubit pure state $\ket{\psi}=a\ket{HH}+b\ket{HV}+c\ket{VH}+d\ket{VV}$ can
always be written using only two terms:$\ket{\psi}=\alpha\ket{\chi\xi}+\beta\ket{\chi^\perp\xi^\perp}$,
where $\ket{\chi}$ ($\ket{\xi}$) is orthogonal to $\ket{\chi^\perp}$ ($\ket{\xi^\perp}$), and $\alpha$ and $\beta$ 
satisfy $|\alpha|^2+|\beta|^2=1$.

Now we describe how to prepare such a state.
The creation of the entangled state 
\begin{equation}
\label{eqn:DC}
\cos\theta |H_{A}H_{B}\rangle+ e^{i\phi}\sin\theta|V_{A}V_{B}\rangle
\end{equation}
 from two-crystal downconversion
was proposed in Ref.~\cite{Kwiat_Waks_White_Appelbaum_Eberhard99}. 
Consider two identically cut thin nonlinear crystals. Suppose the first crystal's
optic axis lies in the vertical plane defined by the directions of pump beam
and the vertical polarization. Assuming type-I phase matching, a $V$-polarized pump will produce
 two $H$-polarized daughter photons. We denote this process by
$\ket{V}\rightarrow \ket{H_A}\otimes\ket{H_B}$.  
If the pump is $H$-polarized,
no downconversion process will take place. Suppose the second crystal is
placed at an orientation rotated from the first crystal by $90^\circ$
about the pump direction. 
An $H$-polarized pump will now produce a pair of $V$-polarized photons
$\ket{H}\rightarrow \ket{V_A}\otimes\ket{V_B}$, 
whereas no downconversion will occur if the pump 
is $V$-polarized~\cite{Migdall97}.
With the two crystals placed in contact with each other, a pump in the state
$\cos\theta |V\rangle+ e^{i\phi}\sin\theta|H\rangle$ will produce
a pair of photons in the state
\begin{equation}
\nonumber
\cos\theta |H_{A}H_{B}\rangle+ e^{i\phi}\sin\theta|V_{A}V_{B}\rangle,
\end{equation}
where $\theta$ and
$\phi$ are tuned using waveplates acting on the pump 
polarization ~\cite{Kwiat_Waks_White_Appelbaum_Eberhard99}.
($\phi$ can also be tuned with, e.g., a variable waveplate 
acting on just one of the downconversion photons.)

Choosing the local unitary transformations $\hat{U}_A$ and $\hat{U}_B$
such that
\begin{subequations}
\begin{eqnarray} 
\hat{U}_A\{\ket{H},\ket{V}\}& \rightarrow& \{\ket{\chi},\ket{\chi^\perp}\},\\
\hat{U}_B\{\ket{H},\ket{V}\}& \rightarrow& \{\ket{\xi},\ket{\xi^\perp}\},
\end{eqnarray}
\end{subequations}
we can achieve the arbitrary two-qubit pure state $\ket{\psi}$
by starting with an entangled state with $(\cos\theta,e^{i\phi}\sin\theta)=(\alpha,\beta)$ [up to an irrelevant overall
phase], followed by the corresponding local rotations $\hat{U}_A$ and $\hat{U}_B$
\begin{eqnarray}
\label{eqn:abcd}
\!\!\!\!\!\!\!&&\hat{U}_A\otimes \hat{U}_B\big(\cos\theta |H_AH_B\rangle+ e^{i\phi}\sin\theta|V_AV_B\rangle\big)
\nonumber \\
\!\!\!\!\!\!\!&&=
a\ket{H_AH_B}+b\ket{H_AV_B}+c\ket{V_AH_B}+d\ket{V_AV_B}.
\end{eqnarray}
The two rotations can be obtained in the process of Schmidt decomposing
$\ket{\psi}$~\cite{NielsenChuang00}; see also Appendix~\ref{sec:ArbitraryPure}
for an explicit construction of the appropriate $\hat{U}$'s, $\alpha$, and $\beta$ given $\{a,b,c,d\}$. 

In practice, any $SU(2)$ rotation such as $\hat{U}_A$ and $\hat{U}_B$ on a polarization state can be implemented by
combinations of half- and quarter-waveplates~\cite{Nick03}---preferably 
zero-order waveplates~\cite{ZeroOrder}, for which 
the retardance is barely sensitive to deviation from the central frequency.
That is to say, the action of waveplates, $\hat{U}$, can be assumed to
be $\epsilon$-independent (at least in the frequency range set by
the interference filter before detection), i.e.,
\begin{equation}
\hat{U}\big\{|\chi_j\rangle\otimes\!\!\int\!\! d\epsilon A(\epsilon)|\frac{\omega}{2} \pm \epsilon\rangle\big\}
\approx \sum_k 
U_{kj}|\chi_k\rangle\otimes\!\!\int\!\! d\epsilon A(\epsilon)|\frac{\omega}{2} \pm \epsilon\rangle,
\end{equation}
where $U_{kj}$ are the elements of a unitary 
matrix that is independent of $\epsilon$. We shall assume throughout this paper that unitary transformations by waveplates are ideal and independent of deviation from the central frequency.

\subsection{Via interferometry}
\label{sec:ArbPureSch2}
\begin{figure} 
\vspace{0.5cm}
\centerline{
\rotatebox{0}{
        \epsfxsize=8.5cm
        \epsfbox{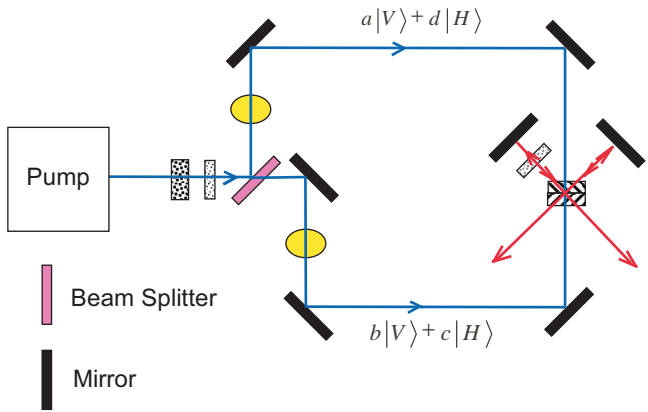}
}
}
\caption{(Color online) Arbitrary pure states via inteferometry.  }
\label{fig:ArbPureSch2}
\end{figure}
A second 
method for creating arbitrary pure states is shown in 
Fig.~\ref{fig:ArbPureSch2} and is a modification from the setup
of Ref.~\cite{HerzogKwiatWeinfurterZeilinger95}.  As discussed in Sec.~\ref{sec:pure}, 
via SPDC (assuming type-I phase matching), a pump in a polarization state
$\alpha\ket{H}+\beta\ket{V}$ will generate an entangled photon pair in the 
state (up to some irrelevant phases)
$\alpha\ket{VV}+\beta\ket{HH}$.
With a half-waveplate, this entangled state can be further transformed into
$\alpha\ket{VH}+\beta\ket{HV}$.
Now, an arbitrary pure two-photon polarization state
$a\ket{HH}+b\ket{HV}+c\ket{VH}+d\ket{VV}$ 
can be regarded as a superposition of two 
(un-normalized) parts: 
$a\ket{HH}+d\ket{VV}$ and   
$b\ket{HV}+c\ket{VH}$. 
The first part can be created from 
the (un-normalized) pump state 
$\ket{\psi_U}\equiv a\ket{V}+d\ket{H}$.
To create the second part, we need the 
(un-normalized) pump state 
$\ket{\psi_L}\equiv b\ket{V}+c\ket{H}$, 
from which SPDC yields the two-photon state 
$b\ket{HH}+c\ket{VV}$. 
Again, a half-waveplate in one arm (or equivalently, passing through a
quarter-waveplate twice) can transform this state into $b\ket{HV}+c\ket{VH}$.
By coherently superposing the above two processes, as shown in
Fig.~\ref{fig:ArbPureSch2}, the fully arbitrary pure two-qubit state
$a\ket{HH}+b\ket{HV}+c\ket{VH}+d\ket{VV}$ can be created.  
The amplitude of each process, which is determined by the relative
values of $\ipr{\psi_U}{\psi_U}$ and $\ipr{\psi_L}{\psi_L}$,
can be adjusted by the transmission through the beam splitter.
Moreover, coherent superposition can be achieved by balancing
the two path lengths. Thus, by combining a well-balanced interferometer 
and the process of spontaneous downconversion we can realize arbitrary 
two-photon polarization {\it pure\/} states. In the next section we shall describe two schemes capable of producing 
arbitrary two-photon polarization {\it mixed\/} states. 

\begin{table*}[t]
\begin{center}
\begin{tabular}{|r|c|c|c|c|l|l|}
\hline  
\begin{tabular}{l}
{\rm Schemes}
\end{tabular}
   & 
    \begin{tabular}{l}
    Synthesizable \\ states
    \end{tabular}
    &CP
   &\begin{tabular}{l}
   {\rm NLC}
   \end{tabular}  & 
   \begin{tabular}{l}{\rm Other} \\{\rm Optics}\end{tabular}
   &Advantages  &
  {\rm Disadvantages}
   \\\hline
 {I} : Fig.~\ref{fig:ArbMixedSch1} & \begin{tabular}{l}
 Arbitrary \\
 two qubits 
 \end{tabular}
 & 15& 8 & $38$ & 
 Arbitrary states
   &\begin{tabular}{l}
  1. Birefringence of crystals causes additional \\  
  \phantom{1.} rotations and possible decoherence\\
   2. Requires precise spatial-mode alignment \\
   3. Narrow opening angles of
    downconversion \\
    \phantom{3.} require long path difference for mixing\\
    4. Potential loss of downconverted photons \\
    5. Waveplate imperfection and wedges, esp. \\
       \phantom{5.} at  early stages, cause 
   beam deviation 
   \end{tabular}
   \\ \hline
 {II} : Fig.~\ref{fig:ArbMixedSch2} 
 &  \begin{tabular}{l}
 Arbitrary \\
 two qubits 
 \end{tabular}
 &15&2 & $48^\dagger$ 
   &
   \begin{tabular}{l} 
   1. Arbitrary states \\
   2. Not lossy \\
   \phantom{2.} in downconversion\\
   3. Only two crystals
   \end{tabular}& 
\begin{tabular}{l}
   1. Requires interferometer stablization  \\
   2. Need to compensate reflection-induced \\
   \phantom{2.} transformations from mirrors \\
   3. Variable beam splitters difficult to tune \\
   4. Lossy in pump
   \end{tabular} \\ \hline
 { III}: Fig.~\ref{fig:RedMixedSch1} &
  \begin{tabular}{l} MEMS Eq.~(\ref{eqn:MEMS}),\\
                 Werner~(\ref{eqn:Werner}), \\  Collins-Gisin~(\ref{eqn:CG})\\ and states~(\ref{eqn:D1})
		 \end{tabular} 
&$\ge10$  & 2 & $10^\ddagger$
		 
   &
    \begin{tabular}{l} 
   1. Partially tested \\
   \phantom{1. }\cite{JoeCG,NickKwiat,Altepeter}\\
   2. Minimal spatial-  \\
   \phantom{2.} mode matching \\
   3. Only two crystals
   \end{tabular}
   &
		  \begin{tabular}{l}
1. Probably not arbitrary states\\
2. No complete theory for more than  one\\
\phantom{2. }decoherer per arm
   \end{tabular}\\ \hline
 {IV}: Fig.~\ref{fig:RedMixedSch2} 
 & \begin{tabular}{l}
          States from  III, \\
	  Bell-diagonal \\
	  states~(\ref{eqn:BellD})\\
	  and states~(\ref{eqn:LS})
	   \end{tabular}
	&$\ge12$    &4 & 26 
   &
    \begin{tabular}{l} 
    More states than III \\
   \end{tabular}
   & \begin{tabular}{l}
  1. Probably not arbitrary states \\
  2. Birefringence of crystals causes additional \\  
  \phantom{1.} rotations and possible decoherence\\
  3. Requires precise spatial-mode alignment\\
   \end{tabular}\\ \hline
	   
\end{tabular}
\end{center}
\vspace{0cm}
\caption{\label{tbl:comparison}Comparison of the four mixed-state schemes.
CP stands for controllable parameters (out of 15 in total).
The nonlinear crystals (NLC) are used in the downconversion process.
By ``other resources'', we include waveplates (where a general unitary requires, e.g.,
1 half-waveplate and 2 quarter-waveplates, hence counted as 3 elements), mirrors, attenuators, prisms, and decoherers, and we assume that the pump is already polarized. 
The crystal and resource numbers given are sufficient to produce all states
given in the final column. This resource accounting is intended to indicate
the relative complexity of the various schemes; however, the numbers listed
may be reduced for certain states, or possibly by using clever combinations
of elements (e.g., reflections which modify polarization).
${}^\dagger$The resource number listed for Scheme~II is several items lower than a direct counting
from Fig.~\ref{fig:ArbMixedSch2}, which was shown for clarity with extra mirrors. ${}^\ddagger$The resource number listed
for Scheme~III is counted without pump decoherence and with only a single
stage of decoherence, and is thus less than a direct counting from Fig.~\ref{fig:RedMixedSch1}, but is sufficient to synthesize the states
listed. } 
\vspace{0cm}
\end{table*} 

\section{Schemes for arbitrary two-photon polarization mixed states}
\label{sec:Arbmixed}
Any two-qubit mixed
state can be canonically decomposed as follows~\cite{NielsenChuang00}:
\begin{equation}
\label{eqn:canonical}
\rho=\sum_{i=1}^4 \lambda_i\ketbra{\psi_i}, 
\end{equation}
where $\{\ket{\psi_i}\}$ are orthonormal eigenstates of $\rho$.
It is therefore natural to realize $\rho$ simply by mixing its eigenstates
with probabilities proportional to their eigenvalues $\lambda_i$.
As we can synthesize arbitrary pure states from one set of crystals,
individual synthesis of each $\ket{\psi_i}$ is straightforward.
 
\begin{figure} 
\vspace{0.5cm}
\centerline{
\rotatebox{0}{
        \epsfxsize=8.5cm
        \epsfbox{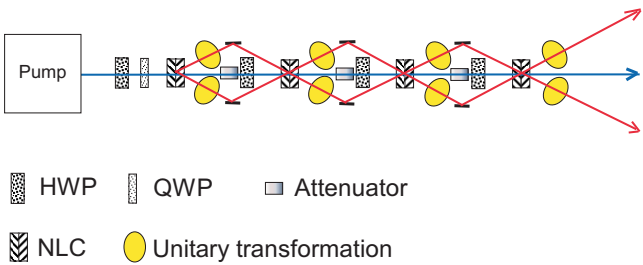}
}
}
\caption{(Color online) Scheme~I employs four sets of nonlinear crystals. The two-photon state
created at the $i$-th set of crystals is chosen such that it is the correct state $\ket{\psi_i}$ after propagation through the subsequent elements.
The necessary local unitary 
transformations at each downconversion location can be readily calculated~\cite{FN:unitary}.   
PBS: polarizing beam splitter; 
HWP: half-waveplate; 
QWP: quarter-waveplate; 
NLC: nonlinear crystals.}
\label{fig:ArbMixedSch1}
\end{figure}
\subsection{Scheme~I: Arbitrary two-qubit mixed states I} 
The first mixed-state scheme is shown in Fig.~\ref{fig:ArbMixedSch1}. 
We have four
pairs of nonlinear crystals, each generating a
pure state that, when propagating to the output, arrives as $\ket{\psi_i}$
~\cite{FN:unitary}.
There is an attenuator in front of each set of crystals (except the first set)
such that the pump intensity $I_i$ going into the $i$-th set of crystals is
proportional to $\lambda_i$ (arranged in decreasing order: $\lambda_1\ge\lambda_2\ge\lambda_3\ge\lambda_4$). 
It is less favorable to attenuate the four downconversion pure states to tune 
the probabilities according to their eigenvalues, 
because direct attenuation of the 
downconversion photons would, 
in general, result in unpaired photons, i.e., one of the photons would be 
absorbed, but not the other~\cite{Zhang04}.

For a pulsed pump, the mixing is incoherent, 
as the arrival time of the downconversion pair
(relative to the pump pulse) can, in principle, reveal information on  
where the pair was generated. For a CW (continuous wave) pump, one can 
add a path delay (much greater than the pump coherence length~\cite{footnote:coherencelength}), to each pair
such that pair-generation amplitudes at
all sets of crystals are no longer coherent with one another.  We can thus
synthesize $\rho$ by incoherently mixing its eigenstates with appropriate weights.
As the downconversion process is much more likely to produce one rather than multiple pairs 
[e.g., see Eq.~(\ref{eqn:2MS})], multiple pairs can be ignored.

\subsection{Scheme II: Arbitrary two-qubit mixed states II}
\begin{figure}
\centerline{
\rotatebox{0}{
        \epsfxsize=7.8cm
        \epsfbox{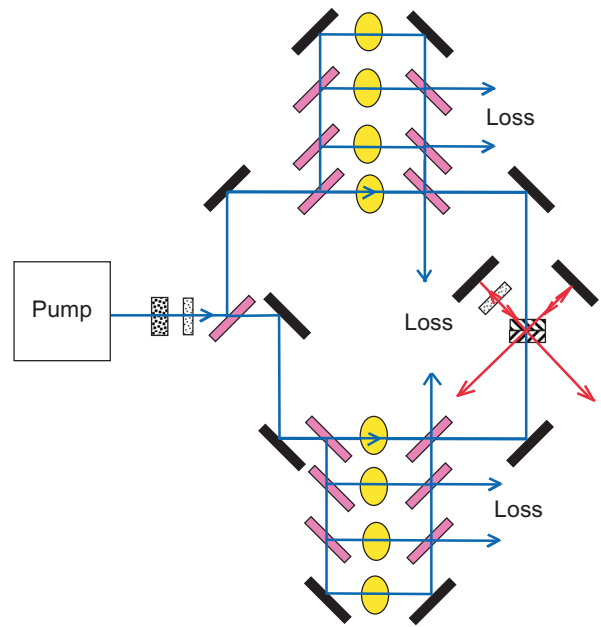}
	}}
\caption{(Color online) Arbitrary two-qubit mixed-state synthesis Scheme~II. 
 The transmission probabilities of the various beam splitters
depend on the desired final state.
The variable beam splitters immediate preceding the unitary rotations
could also be realized by polarizing beam splitters with suitable polarization rotations before and after.}
\label{fig:ArbMixedSch2}
\end{figure}

The interferometric scheme of Sec.~\ref{sec:ArbPureSch2} can also be 
extended to create arbitrary mixed states. 
The full scheme is shown in Fig.~\ref{fig:ArbMixedSch2}. The coherent superposition method
of Fig.~\ref{fig:ArbPureSch2} is used to create each of the four pure states $\ket{\psi_i}$ in the decomposition~(\ref{eqn:canonical}) 
and mix them incoherently in proportion to their eigenvalues $\lambda_i$, 
as in Scheme~I.  Arbitrary weights of mixing can be achieved by controlling
the transmissions of the beam splitters.  In order to mix the four parts 
incoherently, we first use timing information~\cite{Gisin}
such that the state of the pump is 
\begin{equation}
\label{eqn:psip}
\ket{\psi_p}= \sum_{i=1}^4(\ket{\psi_{Ui}}+\ket{\psi_{Li}})\otimes\ket{i}_T.
\end{equation} 
Here, $\ket{\psi_{Ui}}$ and $\ket{\psi_{Li}}$ (both un-normalized) are the 
two parts of the pump state that will, ultimately, yield the corresponding 
pure state $\ket{\psi_i}$~\cite{FN:probabilities}.  
The factors $\ket{i}_T$ ($i=1,\ldots,4$) 
encode timing information; there is no coherence between paths 
labelled by distinct values of $i$, i.e., $\ipr{i}{j}_T=\delta_{ij}$. 
For this absence of coherence to hold, the path-length difference 
between any two upper (or lower) unmatched paths must be greater than 
the pump coherence length~\cite{footnote:pathlength}. As long as coherence is maintained for 
the corresponding pairs of states, 
$\ket{\psi_{Ui}}$ and $\ket{\psi_{Li}}$ (for $i=1,\ldots,4$),
but the time differences for the $i$'s are distinguishable, 
the output state is the desired mixed state, once timing information 
is traced over, i.e., discarded~\cite{Purification2}. 
Note, however, that there is no difference in
{\it relative\/} timing between signal and idler photons. The timing
information is coupled solely to the pump photons; because this
timing information is traced over (ignored), downconversion
produces an incoherent mixture of four two-photon states.
Also note that with a CW pump, the only possibility to detect
any coherence in the timing information would be to include
similar unbalanced interferometers in the downconversion output.

The difference between the 
present scheme (II) and the previous one (I) is that for Scheme~II each of the four pure 
states is created probabilistically in the same downconversion crystals, 
whereas for Scheme~I each of the four pure states is created in downconversion 
crystals at distinct locations.  Both schemes yield arbitrary two-qubit 
mixed states by incoherent temporal mixing. 

\section{Resource-optimized schemes for mixed states}
\label{sec:Redmixed}
In this section we describe two reduced schemes (III and IV) that provide
practical ways to realize several important families of states that 
are currently of interest. Scheme III, whose feasibility has been
demonstrated experimentally, emerges as an effort to reduce
the number of downconversion crystals to two by 
sacrificing the generality of the synthesizable
states. Scheme~IV further extends
the set of synthesizable states by employing two sets of crystals
and the mixing technique introduced in Scheme~I.

\subsection{Scheme~III: Filling the tangle-entropy plane}
Recall that Scheme I requires the use of, at most, four sets of SPDC crystals. 
Since fewer crystal sets would be more economical and likely easier 
to implement, we thus propose a modified scheme, which uses only one set 
of SPDC crystals but relies on ``controlled'' decoherence.  Although we 
do not yet know whether this scheme can generate {\it arbitrary\/} 
two-qubit states, it can synthesize several important families of mixed 
states, including states with all physically allowed values of
entanglement (characterized, e.g., by ``tangle'') and 
mixedness (characterized, e.g., by the linear entropy~\cite{TangleEntropy}).

\begin{figure}
\centerline{
\rotatebox{0}{
        \epsfxsize=7cm
        \epsfbox{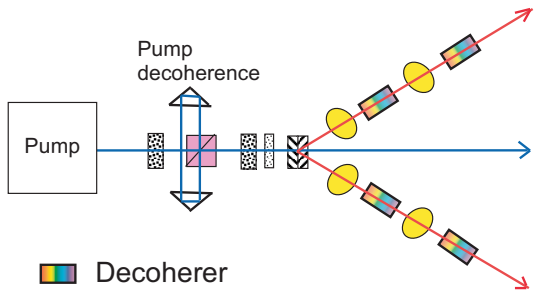}
	}}
\caption{(Color online) Scheme~III employs decoherence. Downconversion photon pairs
can be decohered, as well as pump photons. The decoherers are thick birefringent
crystals, which separate different polarizations and decrease the coherence
between them. Also shown is a possible decoherence on the pump beam:
the vertical polarization component experiences an adjustable extra delay.}
\label{fig:RedMixedSch1}
\end{figure}

We use thick birefringent crystals with thickness $L$ 
as \lq\lq decoherers\rlap.\rq\rq\thinspace\ 
Their effect on a polarization state of definite frequency $\omega$ is (see, e.g.,~\cite{Berglund00,Nick03})
\begin{equation}
{\rm D}(L)|\chi_j\rangle\otimes|\omega\rangle=e^{i n_j L\omega/c}|\chi_j\rangle\otimes|\omega\rangle,
\end{equation}
where the optic axis is assumed to be along, say,
$\chi_2$, i.e., the $V$ direction, and $n_j$ is the refractive
index for the $j$-th polarization state. The decohering elements entangle
the polarization and frequency degrees of freedom.  
In the output, only polarizations are detected, so
we have to trace over the frequency degree of freedom in the joint pure
state (of polarizations and frequencies).  In general, we are then left 
with a {\it mixed\/} two-photon polarization state.
In the present scheme, we can have several decoherers 
in each arm, along with arbitrary unitary rotations between the decoherers 
(only two are shown in Fig.~\ref{fig:RedMixedSch1}). 

In addition to directly decohering the downconversion photons, 
one can also decohere the pump photons before downconversion, 
as indicated in Fig.~\ref{fig:RedMixedSch1}. However, as mentioned
previously, the pump typically has a much longer coherence length than the 
downconversion photons do, and hence may require much greater relative
birefringent delays, e.g., unbalanced polarization interferometers, 
 to achieve decoherence. 

In the limit we are considering, i.e., 
$l_p\gg |\Delta n| L_{1/2} \gg l_{si}$ 
(where $\Delta n\equiv(n_{\rm V}-n_{\rm H})$, 
and $L_1$ and $L_2$ are respective thicknesses of the decoherers),   
decohering the pump in addition to the downconversion photons does 
not provide further control beyond simply decohering the downconversion 
photons. Hence, in the following analysis we shall not consider 
decohering the pump.

Consider a pure initial polarization state of downconversion 
pairs~(\ref{eqn:abcd}):
 $\ket{\psi^{(1)}}=a\ket{HH}+b\ket{HV}+c\ket{VH}+d\ket{VV}$
(which is created by the method described in Sec.~\ref{sec:pure}). 
After the decoherers $D(L_1)$ and $D(L_2)$, one in each
of the two arms, the state is 
\begin{equation}
\ket{\psi}=D(L_1)\otimes D(L_2)\ket{\psi^{(1)}}.
\end{equation}
Tracing over the frequencies~\cite{FootnoteTraceF}, 
the reduced density matrix for the polarization state is
(with $\rho_\psi\equiv \ket{\psi}\bra{\psi}$)
\begin{equation}
\rho={\rm Tr}_\epsilon\,\rho_\psi
=\!\!\!\int \!d\epsilon' \bra{\frac{\omega}{2}+\epsilon',\frac{\omega}{2}-\epsilon'}
\rho_\psi 
\ket{\frac{\omega}{2}+\epsilon',\frac{\omega}{2}-\epsilon'}.
\end{equation}
In the limit $L_1,L_2\gg c/(\delta_\epsilon |\Delta n|)$, where 
$\Delta n$ is assumed to be independent of $\epsilon$, 
the resulting polarization mixed state is (in the $\left\{\ket{HH},\ket{HV},\ket{VH},\ket{VV}\right\}$ basis)
\begin{equation}
\label{eqn:D1}
\begin{pmatrix}
|a|^2 & 0 & 0 & fad^* \cr
0 & |b|^2 & 0 & 0 \cr
0 & 0 & |c|^2 & 0 \cr
f^* a^*d & 0 & 0 & |d|^2 
\end{pmatrix},
\end{equation} 
where $f$ is a complex function of $L_1$ and $L_2$ whose exact form depends on
$A_{si}(\epsilon)$. 
For $A_{si}(\epsilon)$ of gaussian form, as in Eq.~(\ref{eqn:Asi}), 
$f$ is given by 
\begin{equation}
f=\exp{\left(-\frac{1}{2}\left[\frac{\Delta n(L_1-L_2)}{c/\delta_\epsilon}\right]^2 \right)}
e^{-i{\Delta n(L_1+L_2)\omega}/{2c}}.
\end{equation}
Note that $|f|\le 1$, with $|f|=1$ for $L_1=L_2$.

The family of two-qubit mixed states described by (\ref{eqn:D1}) 
turns out to be of the form hypothesized by 
Munro {\it et al.\/}~\cite{Munro_James_White_Kwiat01} in their search for
the maximally entangled mixed states (MEMS), which define
the boundary of physically allowed states
on the tangle-entropy plane~\cite{WhiteJamesMunroKwiat02,Munro_James_White_Kwiat01}.  
(The family~(\ref{eqn:D1}) also contains other maximally entangled mixed 
states, corresponding to different charaterizations of  entanglement and entropy ~\cite{WeiNemotoGoldbartKwiatMunroVerstraete03}.)\thinspace\ 
Although states in this family actually fill the physically allowed region of  
the tangle-entropy plane, this does not mean that the family contains all two-qubit mixed states. 
In fact, the family~(\ref{eqn:D1}) has only 4 independent real parameters 
excluding the phase of the off-diagonal element. By including the 6 additional
real parameters coming from the two arbitrary local $SU(2)$ transformations, 
we can thus control 10 of the 15 real parameters associated with general 
two-qubit mixed states. This assumes a {\it single\/} decoherer in each arm.
The full capability of Scheme~III with an arbitrary number of 
decoherence stages is a difficult theoretical problem that requires further investigation.

Next we specifically describe how to generate maximally entangled
mixed states, Werner states, and a particular class of
mixed states recently discussed by Collins and Gisin~\cite{CollinsGisin03}.
The maximally entangled mixed states found by 
Munro {\it et al.}~\cite{Munro_James_White_Kwiat01} are of the form 
\begin{subequations}
\begin{eqnarray}
\rho_{\rm MEMS}=
\begin{cases} \ \rho_{\rm I}(\mygamma), &{\rm for \ }\frac{2}{3}\le\mygamma\le 1;\cr
              \ \rho_{\rm I\!I}(\mygamma), &{\rm for\ } 0\le\mygamma\le\frac{2}{3};\cr
            \end{cases}\quad\quad&& \\
\rho_{\rm I}(\mygamma)\!=\!\begin{pmatrix}\frac{\mygamma}{2} \!&\! 0 \!&\! 0 & \frac{\mygamma}{2} \cr
             0  \!&\! 1\!\!-\!\!\mygamma \!&\! 0 & 0 \cr
             0  \!&\! 0 \!&\! 0 & 0 \cr
             \frac{\mygamma}{2} \!&\! 0 \!&\! 0 & \frac{\mygamma}{2}\cr
             \end{pmatrix}\!, \ \rho_{\rm I\!I}(\mygamma)\!=\!\begin{pmatrix}
 \frac{1}{3} & 0 & 0 & \frac{\mygamma}{2} \cr
             0  & \frac{1}{3}& 0 & 0 \cr
             0  & 0 & 0 & 0 \cr
             \frac{\mygamma}{2} & 0 & 0 & \frac{1}{3}\cr
             \end{pmatrix}\!.&&
\end{eqnarray}
\label{eqn:MEMS}
\end{subequations}
Here, an irrelevant phase
in the nonzero off-diagonal elements has been set to zero.
For $\rho_{\rm I}(\mygamma)$, we only need to generate
a pure state of the form
\begin{equation}
\sqrt{\frac{\mygamma}{2}}\ket{HH}+\sqrt{1-\mygamma}\ket{HV}
+\sqrt{\frac{\mygamma}{2}}\ket{VV},
\end{equation}
followed by decoherers with thicknesses $L_1=L_2$. 
For $\rho_{\rm II}(\mygamma)$, we start with
\begin{equation}
\sqrt{\frac{1}{3}}\ket{HH}+\sqrt{\frac{1}{3}}\ket{HV}
+\sqrt{\frac{1}{3}}\ket{VV},
\end{equation}
followed by decoherers with thicknesses $L_1$ and $L_2$ such that
$|f(L_1,L_2)|=3\mygamma/2$~\cite{NickKwiat}. This requires either prior
knowledge of $A_{si}(\epsilon)$ or the tuning of $(L_1-L_2)$ so as to
obtain the correct reduction factor $|f|$. 
Similarly, to prepare the Werner states of the form
\begin{eqnarray}
\rho_{\rm W}(r)  &\equiv&\mygamma|\Phi^+\rangle\langle\Phi^+|\!+\!\frac{1-\mygamma}{4}\openone
\nonumber \\
&=&
       \begin{pmatrix}
       \frac{1+\mygamma}{4} & 0 & 0 & \frac{\mygamma}{2} \cr
               0  & \frac{1-\mygamma}{4} & 0 & 0 \cr
               0  & 0 &  \frac{1-\mygamma}{4} & 0 \cr
               \frac{\mygamma}{2} & 0 & 0 & \frac{1+\mygamma}{4}
               \end{pmatrix},
\label{eqn:Werner}
\end{eqnarray}
[with $\ket{\Phi^+}\equiv(\ket{HH}+\ket{VV})/\sqrt{2}$\,], 
we start with the pure state
\begin{equation}
\sqrt{\frac{1\!+\!\mygamma}{4}}\ket{HH}+
\sqrt{\frac{1\!-\!\mygamma}{4}}\ket{HV}+
\sqrt{\frac{1\!-\!\mygamma}{4}}\ket{VH}+
\sqrt{\frac{1\!+\!\mygamma}{4}}\ket{VV},
\end{equation}
and follow with decoherers with thicknesses $L_1$ and $L_2$ such that
$|f(L_1,L_2)|=2\mygamma/(1+\mygamma)$.  Analogous procedures yield 
the other forms of the Werner states, i.e., with other maximally
entangled components.

Using these methods, several maximally entangled mixed states and 
Werner states have been synthesized experimentally, with high 
fidelities~\cite{Fidelity} between the experimentally produced 
states and the theoretical target states.  For example, the MEMS 
\begin{equation}
\begin{pmatrix}
\frac{1}{3} & 0 & 0 & \frac{1}{3}\cr
0 & \frac{1}{3} & 0 & 0 \cr
0 & 0 & 0 & 0\cr
\frac{1}{3}& 0 & 0 & \frac{1}{3}
\end{pmatrix}
\end{equation}
was used to investigate entanglement purification protocols~\cite{NickKwiat},
and the separable Werner state 
\begin{equation}
\begin{pmatrix}
\frac{1}{3} & 0 & 0 &\frac{1}{6}\cr
0 & \frac{1}{6} & 0 & 0\cr
0 & 0 & \frac{1}{6} & 0 \cr
\frac{1}{6} & 0 & 0 & \frac{1}{3}
\end{pmatrix}
\end{equation}
was used to perform ancilla-assisted process tomography without entanglement~\cite{Altepeter}.

Next, we turn to the Collins-Gisin states, particular mixtures of two pure
states: 
\begin{eqnarray}
\label{eqn:CG}
\rho_{\rm CG}(\lambda,\theta)&\equiv& \lambda \ketbra{\psi_\theta}+(1-\lambda)\ketbra{HV}  \\
&=&\begin{pmatrix}  
\lambda\cos^2\theta & 0 & 0 & \lambda\cos\theta\sin\theta \cr
             0  & (1-\lambda) & 0 & 0 \cr
             0  & 0 & 0 & 0 \cr
             \lambda\cos\theta\sin\theta& 0 & 0 & \lambda\sin^2\theta\nonumber
	     \end{pmatrix},
\end{eqnarray}
where
$\ket{\psi_\theta}\equiv \cos\theta\ket{HH}+
\sin\theta\ket{VV}$. Collins and Gisin reported a Bell-like inequality
(which they call I3322) that is {\it inequivalent\/} to the usual CHSH-Bell inequality~\cite{CHSH}, 
in that there are states that do not violate CHSH but {\it do \/}
violate I3322~\cite{CollinsGisin03}.  For example, 
the family of states 
$\rho_{\rm CG}(\lambda,\theta)$ exhibit this behavior
for certain ranges of $\lambda$ and $\theta$
where no violations of CHSH occur.
How can we create these Collins-Gisin states?  
In light of the above examples of MEMS and Werner states, 
we see that we only need to 
generate a pure state of the form
\begin{equation}
\sqrt{\lambda}\cos\theta\,\ket{HH}+\sqrt{ 1-\lambda}\,\ket{HV}
+\sqrt{\lambda}\sin\theta\,\ket{VV},
\end{equation}
followed by a decoherence with $L_1=L_2$. Such states
have been experimentally realized and used to study various
tests for entanglement and nonlocality~\cite{JoeCG}.

States described by Eq.~(\ref{eqn:D1}) (plus those derived  from them by local
unitary transformations) are not the most general form that Scheme~II can
achieve. For example, if, via downconversion, we prepare the pure state $\ket{\Psi^+}\equiv(\ket{HV}+\ket{VH})/\sqrt{2}$,
apply decoherers of common thickness $L$ ($\gg l_{si}$) in both arms, 
and then rotate each photon polarization by $45^\circ$, 
followed by a second set of  decoherers with the same thicknesses, we would 
generate a mixed state of the form
\begin{equation}
\begin{pmatrix}
\frac{1}{4} & 0 & 0 &\frac{1}{4}\cr
0 & \frac{1}{4} & \frac{1}{8} & 0\cr
\noalign{\smallskip}
0 & \frac{1}{8} & \frac{1}{4} & 0 \cr
\frac{1}{4} & 0 & 0 & \frac{1}{4}
\end{pmatrix},
\end{equation}
up to some irrelevant phases. This state does not belong to the
family~(\ref{eqn:D1}), obtained with only one stage of decoherence,  
suggesting that using multiple decoherences may enable control over
 {\it more\/} 
than the 10 independent parameters allowed by a single decoherence.  
Further theoretical investigation is needed to determine the most general
states obtainable.

\begin{figure}
\centerline{
\rotatebox{0}{
        \epsfxsize=8.5cm
        \epsfbox{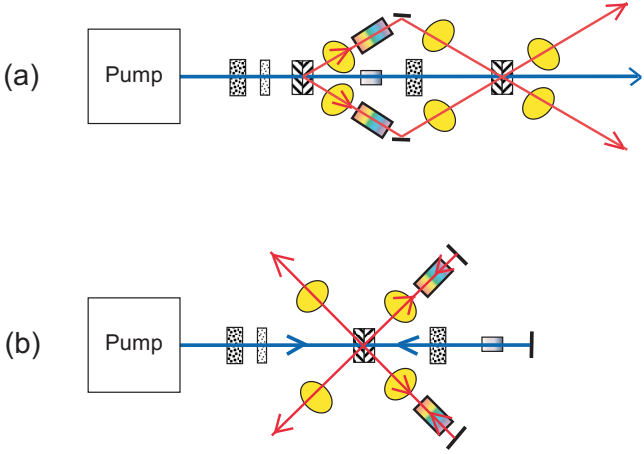}
	}}
\caption{(Color online) Scheme~IV is a hybrid technique. 
(a)~Mixing a pure state with a mixed state.  
The local unitary transformations immediately after the decoherers 
are used to pre-compensate the effect of local unitary transformations 
used afterward to rotate the pure part, and also to undo any effects 
of passing through the nonlinear crystals (c.f.~Fig.~\ref{fig:ArbMixedSch1}). 
(b)~A reduced setup of the method in (a), using only one set of nonlinear 
crystals, and retro-reflecting the pump back through the nonlinear crystals 
and the first photon pair back into the spatial modes of the second pair. 
Setup~(b) is less general than (a), as the pure-state part cannot be chosen
arbitrarily.}
\label{fig:RedMixedSch2}
\end{figure}

\subsection{Scheme~IV: A hybrid technique}
From Scheme~I it appears that one needs four sets of nonlinear crystals in 
order to synthesize fully general rank-four mixed states, whereas 
from Scheme~III one can create rank-four mixed states of the restricted 
form~(\ref{eqn:D1}) (up to local unitary transformations) with
a single set of crystals.  As we now discuss, 
by using a hybrid scheme one can, with only two sets (or in some cases, even 
just a single set) of crystals, generate a larger class of rank-four mixed states. 

The idea is as follows.
  Suppose that the state $\rho$ can be decomposed into 
\begin{equation}
\label{eqn:LS}
\rho=p \,\sigma + (1-p)\ketbra{\psi}. 
\end{equation}
If the mixed state $\sigma$ can be created by Scheme~III [e.g., states in Eq.~\ref{eqn:D1}], we can then mix, 
with appropriate weights, $\sigma$ (as created from a first set of 
crystals) with the pure state $\ketbra{\psi}$ (from a second set), 
and thus obtain $\rho$~\cite{FootnoteAttenutate}; see Fig.~\ref{fig:RedMixedSch2}a.  
Although any two-qubit mixed state always allows a decomposition
into a mixed state plus a pure part~\cite{Lewenstein_Sanpera98}, 
it remains an open question whether there always exists a decomposition 
for which the mixed-state part is achievable by Scheme~III.  
Nevertheless, this hybrid scheme can obviously generate more states than 
Scheme~III alone, since adding the pure part adds more degrees of freedom.

We can make a simple reckoning of the number of parameters of the 
achievable density matrices independently controllable.  
Suppose we restrict the mixed-state part $\sigma$ to be produced by
Scheme~III with only one stage of decoherence, i.e., $\sigma$ is of 
the form~(\ref{eqn:D1}) with all parameters real.
Recall that a general two-qubit pure state can be expressed as
\begin{equation}
\ket{\psi}=U_A\otimes U_B \big(\sqrt{\lambda}\ket{HV}+\sqrt{1-\lambda}\ket{VH}\big),
\end{equation}
where $U_A$ and $U_B$ are local unitary transformations, and $\sqrt{\lambda}$
and $\sqrt{1-\lambda}$ are Schmidt coefficients.
Expressed in the basis in which the pure state $\ket{\psi}$ 
is Schmidt decomposed, 
the mixed state $p\,\sigma+(1-p)\ket{\psi}\bra{\psi}$ appears as 
\begin{eqnarray}
\label{eqn:Count12}
&&p\,U^\dagger_A\otimes U^\dagger_B
\begin{pmatrix}
a^2 & 0 & 0 & fad\,\cr
0 & b^2 & 0 & 0 \cr
0 & 0 & c^2 & 0 \cr
f ad & 0 & 0 & d^2 
\end{pmatrix}U_A\otimes U_B 
\nonumber\\
&&+(1-p)
\begin{pmatrix}
0 & 0 & 0 & 0 \cr
0&\lambda & \sqrt{\lambda(1-\lambda)}&0 \cr
0&\sqrt{\lambda(1-\lambda)}  & 1-\lambda &0\cr
0 & 0 & 0 & 0 \cr 
\end{pmatrix}.
\end{eqnarray}
For fixed $U_A$ and $U_B$, this gives, in general, 6 independent parameters 
$\{a,b,c,f,p,\lambda\}$ [noting that $d$ is not independent of $\{a,b,c\}$].
Barring some coincidence that, for different pairs of 
$\{a,b,c,d,f\}$ and $\{U_A,U_B\}$, gives the same mixed part, we have in total 12 independent parameters, 
after adding 6 from the local unitaries~\cite{Footnote12}.

One important family of states that this scheme can synthesize (and which 
cannot be generated via Scheme~III with only one stage of decoherence) are 
the arbitrary Bell-diagonal mixed states~\cite{BennettBrassardPopescuSchumacherSmolinWootters96}:
\begin{equation}
\label{eqn:BD}
\rho_B\equiv\lambda_1\ketbra{\Phi^{\!+}}+\lambda_2\ketbra{\Phi^{\!-}}+
\lambda_3\ketbra{\Psi^{\!+}}
+\lambda_4\ketbra{\Psi^{\!-}}.
\end{equation}
Expressed in the  
$\{\ket{HH},\ket{HV},\ket{VH},\ket{VV}\}$ basis
\begin{equation}
 \label{eqn:BellD}
 \rho_{B}=\frac{1}{2}\begin{pmatrix}
 \lambda_1+\lambda_2 & 0 & 0 & \lambda_1-\lambda_2\cr
 0 & \lambda_3+\lambda_4 & \lambda_3-\lambda_4 & 0\cr
 0 & \lambda_3-\lambda_4 & \lambda_3+\lambda_4 & 0\cr
 \lambda_1-\lambda_2 & 0 & 0 &\lambda_1+\lambda_2
 \end{pmatrix}.
\end{equation}
Assuming that 
$|\lambda_3-\lambda_4|\le 1/2$ 
(otherwise $|\lambda_1-\lambda_2|\le 1/2$), 
$\rho_B$ can also be decomposed as 
\begin{equation}
\rho_{\rm B}=
(1-|\lambda_3-\lambda_4|)\rho_1 + |\lambda_3-\lambda_4|\,\ketbra{\Psi},
\end{equation} 
where $(1-|\lambda_3-\lambda_4|)\rho_1$ is 
\begin{subequations}
\begin{equation}
\frac{1}{2}\!\!\begin{pmatrix}
\lambda_1\!\!+\!\!\lambda_2 & 0 & 0 &\lambda_1\!\!-\!\!\lambda_2\cr
0 &\lambda_3\!\!+\!\!\lambda_4\!\!-\!\!|\lambda_3\!\!-\!\!\lambda_4|& 0 & 0 \cr
0 & 0 &\lambda_3\!\!+\!\!\lambda_4\!\!-\!\!|\lambda_3\!\!-\!\!\lambda_4| & 0\cr
\lambda_1\!\!-\!\!\lambda_2 & 0 & 0 & \lambda_1\!\!+\!\!\lambda_2
\end{pmatrix}
\end{equation}
and the pure-state part is a Bell state
\begin{equation}
\ket{\Psi}=\frac{1}{\sqrt{2}}\big(\ket{HV}+{\rm
sgn}(\lambda_3\!-\!\lambda_4)\ket{VH}\big).
\end{equation}
\end{subequations}
Here ${\rm sgn}(x)$ is the sign function, which gives a factor of $\pm 1$,
depending on the sign of $x$.
It is clear that $\rho_1$ belongs to the family of states~(\ref{eqn:D1}), 
and hence can be synthesized by Scheme~III with one stage of decoherence; 
on the other hand, $\ket{\Psi}$ is a Bell state, which can be easily generated.  
Furthermore, the weight of $\rho_1$ is not less than that of $\ketbra{\Psi}$, 
so there is no need to attenuate the intensity of the mixed part~\cite{FootnoteAttenutate}.  
Therefore, Scheme IV can synthesize {\it any\/} Bell-diagonal state $\rho_{\rm B}$.
The Bell-diagonal states, if entangled, can be readily distilled via the BBPSSW 
scheme~\cite{BennettBrassardPopescuSchumacherSmolinWootters96,PanSimonBruknerZeilinger01} into states with more entanglement.  They also have the 
property that, for a given set of eigenvalues,  
they achieve the maximal violation of the CHSH-Bell inequality~\cite{VerstraeteWolf02}.  

For certain states this hybrid scheme can also be implemented 
via a single set of crystals, by reflecting the source and the downconversion 
pair back through the crystals with a mirror; see Fig.~\ref{fig:RedMixedSch2}b. 
However, in this case, the pure-state part cannot be arbitrarily chosen, 
as the local unitary transformations needed to create  the mixed part $\sigma$ 
and the pure part $\ket{\Psi}$ are no longer independent. 
The mixed part $\sigma$ is obtained via locally rotating $\alpha_1\ket{HH}+\beta_1\ket{VV}$ by $\hat{U}_A\otimes \hat{U}_B$, followed
by a decoherence. As the photons reflect back from the mirrors,
they experience again the same local unitary transformation $\hat{U}_A\otimes \hat{U}_B$. 
The local unitary transformation at the output port is then chosen 
to eliminate this additional effect (by choosing the inverse of this transformation), thereby fixing $\sigma$ to be of
the form~(\ref{eqn:D1});  the pure state part is consequently limited
to the form $\ket{\psi}=\hat{U}_A^{-1}\otimes \hat{U}_B^{-1}\left(\alpha_2\ket{HH}+\beta_2\ket{VV}\right).$
Any further local unitary transformation will rotate $\sigma$ and
$\ket{\psi}$ together and cannot change this relative
relation.
 
\section{Concluding remarks}
\label{sec:conclude}
We have described two approaches for synthesizing arbitary two-qubit pure
states. Based on these,
we have developed four schemes for synthesizing two-qubit photon polarization 
{\it mixed\/} states.
Scheme~I (Fig.~\ref{fig:ArbMixedSch1}) 
requires several sets of downconversion crystals to create arbitrary two-qubit mixed states.
It would be desirable to experimentally synthesize rank-two mixed states using
this scheme, in order to give a proof-of-principle demonstration.
Scheme~II (Fig.~\ref{fig:ArbMixedSch2}) employs temporal mixing
to achieve decoherence.  It offers a second way to realize arbitrary two-photon 
polarization mixed states, but---significantly---requires  only one set of downconversion crystals, at the cost of
requiring several rather large, phase-stabilized interferometers. 
Scheme~III (Fig.~\ref{fig:RedMixedSch1}) provides control over at least 10 of the independent 
real parameters of two-qubit mixed states, 
and gives access to all physically allowed values of entanglement and entropy.
Furthermore, this scheme has been experimentally implemented to synthesize several interesting families of
mixed states, such as Werner states, maximally entangled mixed states, and
Collins-Gisin states~\cite{Altepeter,NickKwiat,JoeCG}. 
The fourth scheme (Fig.~\ref{fig:RedMixedSch2}), extends the range
of Scheme~III (providing control over 12 mixed-state parameters). 
In particular, this scheme can be used to produce arbitrary Bell-diagonal 
states, which are of interest, e.g., in entanglement distillation~\cite{BennettBrassardPopescuSchumacherSmolinWootters96,PanSimonBruknerZeilinger01} 
and maximal violations of Bell inequalities~\cite{VerstraeteWolf02}. 
Although the full capabilities of Schemes~III and IV are not yet entirely clear,
our analysis shows that these two schemes provide practical methods for 
creating quite general mixed states, many of which were previously not
accessible 
experimentally. 
The four mixed-state schemes are summarized and compared in  Table~\ref{tbl:comparison}, including the respective resources, advantages
and disadvantages for implementation. 

Once one has well-controlled arbitrary two-qubit sources, they will be 
usable for many quantum information processing applications, such
as testing methods of entanglement 
distillation~\cite{BennettBrassardPopescuSchumacherSmolinWootters96,PanSimonBruknerZeilinger01,NickKwiat,YamamotoKoashiOzdemirImoto03}, 
investigating quantum process tomography~\cite{Altepeter,DeMartini03}, 
characterizing quantum gates~\cite{White03}, 
testing violations of Bell-type inequalities~\cite{CHSH,Kwiat_Waks_White_Appelbaum_Eberhard99,BarbieriDeMartiniDiNepiMataloni03,VerstraeteWolf02} 
by mixed states 
(including a relevant two-qubit Bell inequality~\cite{CollinsGisin03} 
that is inequivalent to the CHSH inequality), 
and exploring the vast, previously inaccessible territory of Hilbert space.
\section*{Acknowledgments}
The authors would like to acknowledge useful discussions with 
Bill Munro and Sam Braunstein.
This work was supported by NSF Award No.~EIA01-21568, 
ARDA, the DCI Postdoctoral Research Fellowship Program, and 
by the MURI Center for Photonic Quantum Information Systems (ARO/ARDA program DAAD19-03-1-0199) .

\appendix
\section{Creating arbitrary pure states}
\label{sec:ArbitraryPure}
In this appendix we explain how to create an arbitrary two-qubit pure 
state, characterized by $\{a,b,c,d\}$ of Eq.~(\ref{eqn:abcd}).  
This amounts to establishing adequate local unitary transformations 
($U_A$ and $U_B$) and post-SPDC pure states Eq.~(\ref{eqn:DC}) of the 
form that SPDC naturally yields.  For convenience, we exchange the 
coefficients $\cos\theta$ for $\alpha$ and $e^{i\phi}\sin\theta$ for $\beta$. 

 To find the appropriate settings and local unitary transformations,
 we need to solve
\begin{eqnarray}
U_A\otimes U_B (\alpha\ket{HH}+\beta\ket{VV})= \nonumber \\a\ket{HH}+b\ket{HV}+c\ket{VH}+d\ket{VV},
\end{eqnarray} 
for $\{U_A,U_B,\alpha,\beta\}$, 
given $\{a,b,c,d\}$ that are properly normalized.  
This equation can be solved either by Schmidt decomposition or
by direct algebraic manipulation.  However, the solution is not unique.

When $ad-bc=0$, the state to synthesize is a product state, which can
be created from an initial state $\ket{HH}$ followed by independent local rotations 
(see, e.g., Ref.~\cite{Nick03}).  For 
\begin{equation}
ad-bc\ne 0,\qquad\qquad
|ad-bc|\ne 1/2, 
\end{equation}
i.e., the case of non-maximally entangled pure states,
one possible solution is  
\def\Wpm{\phantom{-}}
\begin{subequations}
\begin{eqnarray}
&&\alpha=\sqrt{1-\sqrt{1-4|ad-bc|^2}}/\sqrt{2},\\
&&\beta=(ad-bc)/\alpha,\\
&&U_A=\begin{pmatrix}\Wpm u_1 & v_1 \cr
                    -v_1^* &  u_1^*
		    \end{pmatrix}, \ \ 
U_B=\begin{pmatrix} \Wpm u_2 & v_2 \cr
                    -v_2^* &  u_2^*
		    \end{pmatrix},		   
\end{eqnarray}
\end{subequations}
where
\begin{subequations}
\begin{eqnarray}
\!\!\!\!\!\!\!\!\!\!u_1&\equiv&|z_1|/\sqrt{|z_1|^2+|z_3|^2},
\ \ v_1\equiv z_3 u_1^*/ z_1^*, 
\\
\!\!\!\!\!\!\!\!\!\!u_2&\equiv&z_3^*/(z_1z_3+z_2 z4), 
\ \ v_2\equiv z_2/(z_1z_3+z_2z_4),
\end{eqnarray}
\end{subequations}
with $z_1, z_2, z_3, z_4$ defined via
\begin{subequations}
\begin{eqnarray}
z_1&\equiv&(\Wpm a \alpha^* - d^*\beta)/(|\alpha|^2-|\beta|^2),\\
z_2&\equiv&(\Wpm d^*\alpha-a\beta^*)/(|\alpha|^2-|\beta|^2),\\
z_3&\equiv&(-c^*\alpha-b\beta^*)/(|\alpha|^2-|\beta|^2),\\
z_4&\equiv&(-b\alpha^*-c^*\beta)/(|\alpha|^2-|\beta|^2).
\end{eqnarray}
\end{subequations}

When $|ad-bc|=1/2$ there are three possible cases:\\
(i)~$b=c=0$ and $|a|=|d|=1/\sqrt{2}$;\\ 
(ii)~$a=d=0$ and $|b|=|c|=1/\sqrt{2}$;\\
(iii)~$a,b,c,d\ne 0$, $|a|=|d|$, $|b|=|c|$.\\
Case~(i) is already the form we seek.
In case~(ii), an exchange $H\leftrightarrow V$ 
(e.g.,~by a half-waveplate) will do. 
In case~(iii), one possible solution is
\begin{subequations}
\begin{eqnarray}
e^{i\gamma}&\equiv&a/d^*=-b/c^*,
\label{eq:newphase}
\\
\noalign{\smallskip}
(\alpha,\beta)&=&(e^{i\gamma},1)/\sqrt{2},\\
\noalign{\smallskip}
U_A&=&\frac{1}{\sqrt{2}}\begin{pmatrix}
1&e^{i\gamma}\cr
-e^{-i\gamma}&1 
\end{pmatrix}, 
\\ 
\noalign{\smallskip}
U_B&=&\begin{pmatrix} 
d^*-c&d^*+c\cr
-d-c^*&d-c^*
\end{pmatrix}.	
\end{eqnarray}
\end{subequations}
In fact, this last case includes the previous two cases, if one 
interprets the phase in Eq.~(\ref{eq:newphase}) to be the appropriate 
ratio of the nonzero coefficients.
We remark that the solution presented above is not unique, and that 
experimentally one would implement the one that is most convenient. 


\begin{thebibliography}{99}
\bibitem{Purification}
A purification of a mixed state is a hypothetical extension of the state into
an enlarged Hilbert space formed by adding ancillas such that the extended
state is a pure state in the enlarged space. Upon tracing
over the ancillas, the state reduces to the original mixed state~\cite{NielsenChuang00}.
As a two-qubit state $\rho$ can always be decomposed in
the canonical form 
in Eq.~(\ref{eqn:canonical}), an immediate example of its purification is
$\ket{\psi_P}\equiv\sum_{i=1}^4 \sqrt{\lambda_i}\ket{\psi_i}\otimes{\ket{i}_a}$, where
$\ket{1}_a, \ket{2}_a, \ket{3}_a$, and $\ket{4}_a$ are ancilla states.
For example, the $\ket{i}_a$ could be encoded in 
the two-photon polarization 
states of two additional photons: $\ket{HH}, \ket{HV}, \ket{VH}$, and $\ket{VV}$. If there exist
efficient CNOT gates, creating the four-photon pure state $\ket{\psi_P}$ 
is a straightforward procedure of unitary operations. However, such
efficient entangling gates for photon polarizations are not yet available. 
See also~\cite{FootnoteCNOT}.
\bibitem{footnote:coherencelength}
Typical lasers have coherence lengths between 1cm and 1m. Practically,
downconversion coherence lengths are selected experimentally with
narrow-bandpass ($\Delta\lambda$) interference filters and are given by
$l_{si}\approx \lambda^2/\Delta \lambda$. For example, for 700 nm photons 
with 1 nm (5 nm) bandpass interference filters, the coherence length is roughly
500 $\mu$m (100 $\mu$m).
\bibitem{FN:unitary}
The local unitary transformations at each stage can be calculated
backwards, starting with those at the fourth set of crystals.
The unitary transformations caused by the nonlinear crystals themselves
can be compensated by modifying the local unitary transformations.  
Decoherence effects arising from waveplates and crystals are assumed to be negligible.
\bibitem{FN:probabilities}
The probabilities (i.e.,~the eigenvalues or, 
equivalently, the relative intensities) can be included in the normalization 
of the states $\ket{\psi_{Ui}}$ and $\ket{\psi_{Li}}$.
Note that the actual pump state before downconversion includes 
some amplitudes that will inevitably be lost (out of unused beamsplitter 
ports, as indicated by the subscripts ``loss''), so it will be of the form  
\begin{equation}
\ \ \ \ket{\psi_p}= \sum_{i=1}^4(\ket{\psi_{Ui}}+\ket{\psi_{Li}})\otimes\ket{i}_T\ +
\ket{\psi_{U,\rm loss}}+\ket{\psi_{L,\rm loss}}.\nonumber
\end{equation} 
This loss is of no consequence in practice, as the pump 
contains large numbers of photons, and the creation of a particular two-photon
state will be post-selected by coincidence counts.
\bibitem{footnote:pathlength}
For example, for a pump with a 10 cm coherence length, a path-length
difference of 30 cm is sufficient to eliminate coherence to
1\%, assuming the dependence $\exp[-(\Delta L/l_{p})^2/2]$.
\bibitem{Purification2}
One can argue that the state~(\ref{eqn:psip}) is a purification of
the mixed state, with the ``ancillas'' being the timing information; see also
remarks in~\cite{Purification}.
\bibitem{FootnoteTraceF}
In general, one has to integrate over the two independent frequencies 
$\omega_i$ and $\omega_s$ of the idler and singal photons.
However, as we are considering the limit in which the pump is treated
as monochromatic, there is a constraint: $\omega_i+\omega_s=\omega$.
Hence, the only independent variable to integrate over is the deviation 
$\epsilon$ from half of the pump frequency (for either idler
or signal photon).
\bibitem{FootnoteAttenutate}
If $p\le 1/2$, we may need to attenuate the intensity of the 
photon pair from the first set of crystals. 
Alternatively, to losslessly prepare the state,
we can switch the order of the setup such that the pure state
$\ket{\psi}$ is created in the first set of crystals, and
the mixed state $\sigma$ is created in the second. However,
the photons from the former will then have to be pre-compensated to 
eliminate the effect of the subsequent decoherence.  This can be done 
by using the same decoherers but with their optic axes rotated by $90^\circ$.
\bibitem{Footnote12}
If the pump is decohered before the second set of crystals, the downconversion
pair will be in a mixed state (up to local unitaries)
\begin{equation}
\begin{pmatrix}
0 & 0 & 0 & 0 \cr
0&\lambda & g\sqrt{\lambda(1-\lambda)}&0 \cr
0&g^*\sqrt{\lambda(1-\lambda)}  & 1-\lambda &0\cr
0 & 0 & 0 & 0 \cr 
\end{pmatrix}, \nonumber
\end{equation}
where $g$ represents the degree of decoherence of the pump (c.f., $f$ in Eq.~(\ref{eqn:D1})). It is not clear whether  this additional
``knob'' boosts the number of independent controlled parameters from 12 to 13.
\end{thebibliography}
\end{document}